\documentstyle[twocolumn,aps,prl,epsf,floats]{revtex}
\newcommand{\be}{\begin{equation}}
\newcommand{\ee}{\end{equation}}
\newcommand{\ba}{\begin{eqnarray}}
\newcommand{\ea}{\end{eqnarray}}
\newcommand{\bi}{\begin{itemize}}
\newcommand{\ei}{\end{itemize}}

\def\lsi{\raise0.3ex
\hbox{$<$\kern-0.75em\raise-1.1ex\hbox{$\sim$}}}
\def\gsi{\raise0.3ex
\hbox{$>$\kern-0.75em\raise-1.1ex\hbox{$\sim$}}}

\begin{document}
\twocolumn[\hsize\textwidth\columnwidth\hsize\csname
@twocolumnfalse\endcsname

\draft
\title{Primordial magnetic fields, anomalous isocurvature
fluctuations\\
 and Big Bang nucleosynthesis}
\author{ M. Giovannini$^{(b)}$ and M. E. Shaposhnikov$^{(a)}$}
\address{$^{(a)}${{\it CERN, Theory Division, CH-1211 Geneva 23,
Switzerland}}\\
  $^{(b)}${\it DAMTP, Silver Street, CB3 9EW Cambridge,
United Kingdom}}
\date{August 10, 1997}
\maketitle

\vspace*{-4.0cm}
\noindent
\hfill \mbox{CERN-TH/97-138, hep-ph/9708303}
\vspace*{4.0cm}

\begin{abstract}\noindent
We show that the presence of primordial stochastic (hypercharge)
magnetic fields before the electroweak (EW) phase transition induces
isocurvature fluctuations (baryon number inhomogeneities). Depending on
the details of the magnetic field spectrum and on the particle physics
parameters (such as the strength of the EW phase transition and
electron Yukawa couplings) these fluctuations may survive until the Big
Bang nucleosynthesis (BBN). Their lenghtscale may exceed the neutron
diffusion length at that time, while their magnitude can be so large
that sizable antimatter domains are present. This provides the
possibility of a new type of initial conditions for non-homogeneous BBN
or, from a more conservative point of view, stringent bounds on
primordial magnetic fields.
\end{abstract}

%\vspace*{0.1cm}

%\pacs{PACS numbers: 11.10.Wx, 11.15.Ha, 12.38.Mh}
\vskip1.5pc]

Large scale magnetic fields in diffuse astrophysical plasmas represent
a well established observational fact since few decades. It has been
realized through the years that magnetic fields coherent over diverse
length scales at different epochs may have a variety of quite
interesting phenomenological consequences. Magnetic fields coherent
today over scales of the order of 30 Kpc are measured \cite{1b} and
have an important role in the dynamics of the galaxy, for example, in
confining cosmic rays \cite{2}. Magnetic fields at the nucleosynthesis
epoch, even if not directly observable, could change the reaction and
the expansion rate at that time. The success of the homogeneous and
isotropic BBN  provides then interesting bounds on their
existence \cite{3}.

There are neither compelling theoretical arguments nor motivated
phenomenological constraints which could exclude the existence of
magnetic fields prior to the nucleosynthesis epoch. Moreover, to
explain the origin of the galactic magnetic fields some authors often
invoke the dynamo mechanism which might amplify the primordial ``seed"
magnetic field. It is a challenge to produce large scale seeds, and 
different ideas were aimed at this purpose. The energy scales
involved vary from $\sim 100$ MeV for the QCD phase transition
\cite{sigl} to $\sim 100$ GeV in the case of the EW physics
\cite{4,joyce} and even closer to the Planck energy scale 
for inflation or string cosmology \cite{planck,string}.

The purpose of this Letter is the study of the implications of the
primordial magnetic fields which existed even before the EW scale (i. e. for
temperatures $\gsi 100$ GeV). The origin of these seeds is not
essential for us and consequently we simply assume that they were
generated by some mechanism before the EW phase transition. Our main
point is that these fields produce baryon and lepton number
inhomogeneities (isocurvature fluctuations), which could have an impact
on the standard BBN.

Let us start from some qualitative considerations. A unique property of
``unbroken" U(1) gauge interaction is the absence of mass of its
corresponding vector particle. Static ``magnetic" fields are never
screened (in the absence of monopoles) and thus homogeneous fields can
survive in the plasma for infinite time. Under normal conditions (i.e.
small temperatures and small densities of the different fermionic
charges) the SU(2)$\times$U(1)$_Y$ symmetry is ``broken" down to
U(1)$_{EM}$, the massless field corresponding to U(1)$_{EM}$ is the
ordinary photon and the only long-lived field  in the
plasma is the ordinary magnetic one. At sufficiently high
temperatures $T > T_c$, the SU(2)$\times$U(1)$_Y$ symmetry is
``restored", and non-screened vector modes $Y_\mu$ correspond to the
U(1)$_Y$ hypercharge group. Hence, if primordial fields existed at $T >
T_c$, they did correspond to hypercharge rather that U(1)$_{EM}$. There are
essential differences between the interactions of magnetic fields and
the ones of hyper-magnetic fields with matter. 
The ordinary electro-magnetic field
has a vector-like coupling to the fermions, while the coupling of the
hypercharge fields is chiral. Thus, if hyper-electric ($\vec{{\cal
E}}_{Y}$) and hyper-magnetic ($\vec{{\cal H}}_{Y}$) fields are present
simultaneously, they cause a variation of the fermionic number
according to the anomaly equation, $\partial_\mu j_\mu \sim
\frac{g'^2}{4\pi^2} \vec{{\cal H}}_{Y}\cdot \vec{{\cal E}}_{Y}$ (here
$g'$ the hypercharge gauge coupling constant). Now, the presence of
{\em non-homogeneous} hyper-magnetic fields in the EW plasma with (hyper)
conductivity $\sigma_c$ always implies the existence of a related
electric field, $\vec{{\cal E}}_{Y}\sim \frac{1}{\sigma_c} \vec{\nabla}
\times \vec{{\cal H}}_{Y}$. Since for a general stochastic magnetic
background $\langle(\vec{{\cal H}}_{Y}\cdot \vec{\nabla}\times
\vec{{\cal H}}_{Y})^2\rangle \neq 0$, the non-uniform hyper-magnetic
field must produce baryon and lepton density perturbations because of
the anomaly equation. In what follows we compute the amplitude of isocurvature
fluctuations induced by this mechanism and discuss their physical
relevance.

The starting point of our discussion will be the generalization of the
magneto-hydrodynamics (MHD) equations ( valid for ordinary
electro-magnetic plasmas) to
the case of hyper-magnetic fields with anomalous coupling to the  fermionic
degrees of freedom (see also \cite{joyce}). 
These equations have to be used for $T > T_c$. We are
interested in a slow dynamics and we then assume that most of the particle
reactions are in thermal equilibrium in the expanding Universe (the
list of those include all perturbative strong and weak processes,
strong and EW sphalerons, Yukawa interactions of $\mu$, $\tau$
and $s,~c,~b,~t$ quarks). The particle physics processes crucial for
our purposes are those related to the U(1)$_Y$ anomaly and to the
slowest perturbative reactions with right electron chirality flip (e.g.
$e_R + Higgs \rightarrow e_L + W$). Thus, our variables are the
space-dependent hyper-magnetic and electric fields $\vec{{\cal
H}}_{Y},~\vec{{\cal E}}_{Y} $ and right electron chemical potential
$\mu_R(\vec{x})$. The generalized Maxwell equations in a
Friedmann-Robertson-Walker metric with scale factor $a(\tau)$ are
\begin{eqnarray}
& &\frac{\partial{{\vec{H}}_{Y}}}{\partial\tau} = -\vec{\nabla}
\times {\vec{E}}_{Y}
,~~
\frac{\partial{\vec{E}}_{Y}}{\partial\tau}+ {\vec{J}}_{Y} =  4 \mu a
{\vec{H}}_{Y}+
{\vec{\nabla}}\times{ \vec{H}}_{Y}
\nonumber\\
& &{\vec{\nabla}}\cdot{\vec{H}}_{Y}=0,~~
{\vec{\nabla}}\cdot {\vec{E}}_{Y}=0,~~
{\vec{J}}_{Y}=\sigma {\vec{E}}_{Y},~a(\tau) d\tau = dt
\label{maxwell}
\end{eqnarray}
($\vec{E}_{Y}=a^2
\vec{{\cal E}}_{Y}$; $\vec{H}_{Y}=a^2
\vec{{\cal H}}_{Y}$;  $\vec{J_{Y}}=a^3 \vec{j_{Y}}$; $\sigma=
\sigma_{c} a$).
A new term, proportional the right electron chemical potential, comes
from the anomaly contribution to the effective Lagrangian of hypercharge gauge
fields \cite{Redlich},
\begin{eqnarray}
\delta{{\cal L}_{Y, e_{R}}} &=& \mu \epsilon_{ijk} Y^{ij} Y^{k}~,
\nonumber\\
\mu &=& \frac{g'^2}{4\pi^2} \mu_{R},~~
Y_{\alpha\beta}= \partial_{[\alpha}Y_{\beta]}~.
\label{2}
\end{eqnarray}
Since the EW  plasma conductivity is large, $\sigma_{c}\sim \sigma_0 T$
with $\sigma_{0}\simeq 70~-~100$ \cite{jpt}, the time derivatives of the
electric fields in Eq. (\ref{maxwell}) can be neglected (in the MHD
context this is known as ``resistive'' approximation \cite{biskamp}).
This observations allows to express the induced electric field in
terms of the magnetic one,
\begin{equation}
{\vec{{\cal E}}}_{Y} = \frac{{\vec{j}}_{Y}}{\sigma_{c}}
\simeq \frac{1}{\sigma_{c}}\left(4~\mu~
{\vec{{\cal H}}}_{Y}+
{\vec{\nabla}}\times{ \vec{{\cal H}}}_{Y}\right)~~~,
\label{electric}
\end{equation}
and derive an equation for $\vec{{\cal H}}_{Y}$ only. It is interesting
to note that the presence of the fermionic chemical potential induces
an
electric field parallel to the magnetic one.

The set of Eq. (\ref{maxwell}) has to be supplemented by the kinetic
equation for the right electron chemical potential, which accounts for
anomalous and perturbative non-conservation of the right electron
number:
\begin{eqnarray}
\frac{\partial}{\partial t}\left(\frac{\mu_R}{T}\right)&=&
-\frac{g'^2}{4\pi^2 \sigma_{c}T^3} \frac{783}{88}
{\vec{{\cal H}}}_{Y}\cdot \vec{\nabla}\times
{\vec{{\cal H}}}_{Y} 
\nonumber\\
&-& (\Gamma + \Gamma_{{\cal H}} ) \frac{\mu_R}{T},
\label{kinetic}
\end{eqnarray}
where $\Gamma$ is the chirality changing rate,
\be
\Gamma_{{\cal H}} = \frac{783}{22} \frac{\alpha'^2}{\sigma_{c}\pi^2}
\frac{|{\vec{{\cal H}}}_{Y}|^2}{T^2},~~~\alpha'= \frac{g'^2}{4\pi}~~,
\label{abelrate}
\ee
(the numbers $783/88$ and $783/22$ come from the relationship between
$e_R$ number density and chemical potential \cite{joyce}). An
interesting consequence of Eqs. (\ref{kinetic},\ref{abelrate}) is that
in the presence of non-zero uniform magnetic field the right electron
number is non-conserved (if $\Gamma =0$), even for an {\em abelian}
anomaly (cf. Ref. \cite{krs}).

Now we are ready to compute baryon number fluctuations produced in our
scenario. We notice that
at the temperature of the EW phase transition $\sim 100$ GeV,  $\Gamma t
\gg 1$. Then, since
 reactions with right electron chirality flip are in the thermal
equilibrium, the  adiabatic approximation can be used, and 
from Eq. (\ref{kinetic}) we have
\begin{eqnarray}
\frac{\mu_R}{T} \simeq -\frac{\alpha'}{\pi\sigma_{c}T^3}\frac{783}{88}
\frac{{\vec{{\cal H}}}_{Y}\cdot \vec{\nabla}\times
{\vec{{\cal H}}}_{Y} }{\Gamma +\Gamma_{{\cal H}}}.
\label{fluctsym}
\end{eqnarray}
Clearly, a non-uniform distribution of the right electron chemical
potential induces baryon and lepton number perturbations of the same
order of magnitude. We are not going to write the explicit formulae
since there is an important ``storage" effect which amplifies the
estimates of Eq. (\ref{fluctsym}) by many orders of magnitude.
Equations (\ref{electric},\ref{kinetic},\ref{fluctsym}) imply that
\be
 \vec{{\cal H}}_{Y}\cdot \vec{{\cal E}}_{Y} \simeq
\frac{\Gamma}{\Gamma +\Gamma_{{\cal H}}}\frac{1}{\sigma_c}{\vec{{\cal
H}}}_{Y}\cdot \vec{\nabla}\times
{\vec{{\cal H}}}_{Y}  \neq 0~~.
\label{ffdual}
\ee
Now, the change of Abelian Chern-Simons number is given by the time
integral of (\ref{ffdual}). At the EW phase transition the
hyper-magnetic fields are converted into ordinary magnetic fields. The
latter do not have coupling to the anomaly. Thus the CS number has to be
transformed into fermions according to Eq. (\ref{kinetic}). Inserting
the coefficients, we arrive, from Eq. (\ref{ffdual}) at our main result:
\begin{equation}
\delta\left(\frac{n_{B}}{s}\right)(\vec{x}, t_{c}) =
\frac{\alpha'}{2\pi\sigma_c}\frac{n_f}{s}
\frac{{\vec{{\cal H}}}_{Y}\cdot \vec{\nabla}\times
{\vec{{\cal H}}}_{Y}}{\Gamma +\Gamma_{{\cal H}}}
\frac{\Gamma M_0}{T_c^2}
\label{final}
\end{equation}
($n_B$ and $s$ are the baryon and entropy densities,
$s=\frac{2}{45}\pi^2 N_{eff} T^3$, $N_{eff}$ is the effective number of
massless degrees of freedom [$106.75$ for minimal standard model],
$M_0= M_{pl}/1.66 \sqrt{N_{eff}}\simeq 7.1 \times 10^{17}$ GeV). Notice
that in Eq. (\ref{final}) there is an enhancement by 
a factor $ \sim \Gamma M_0/T_c^2$ arising from the
time integration of the anomaly term.

Some comments are now in order.

(i) For the correctness of Eq. (\ref{final}) the EW phase transition
should be strongly first order. Moreover, a necessary condition for
EW baryogenesis \cite{ms} must be satisfied. In the opposite
case all baryon number fluctuations will be erased by SU(2) sphalerons
as it happens in the minimal standard model (MSM) \cite{msm}, while this
is not necessarily the case for the supersymmetric and other extensions
of the standard model \cite{mssm}.

(ii) Besides the primordial hyper-magnetic field, an essential quantity
which fixes the amplitude of the isocurvature fluctuations is the rate
of perturbative right electron chirality flip, $\Gamma$. For $\Gamma_H
\gsi \Gamma$ the amplitude of baryon number fluctuations {\em does not
depend on the magnitude of the magnetic field fluctuations} and it is
determined just by their spectral slope. For $\Gamma_H \lsi \Gamma$ the
rate of right electron chirality flip cancels out and the isocurvature
fluctuations are fixed both by the magnitude and by the spectral slope
of the primordial magnetic fields. In the MSM the rate $\Gamma$ depends
crucially upon the electron Yukawa coupling and is known to be $\Gamma
= T\frac{T_R}{M_0}$, where $T_R \simeq 80$ TeV is the freezing
temperature \cite{cko} of the right electrons. This number appears to
be too small to allow any interesting fluctuations. However, in the
extensions of the standard model the rate $\Gamma$ is naturally larger
than in MSM. For example, in the MSSM the right-electrons Yukawa
coupling is larger by a factor $1/\cos{\beta}$, which may increase the
value of $T_R$ by 3 orders of magnitude for experimentally allowed
$\tan (\beta) \sim 50$. Cosmologically interesting fluctuations arise
at $T_R > T^*\simeq 10^5$ TeV \cite{massimo}.

We will assume now that $\Gamma \gsi \Gamma_H $, but similar
conclusions hold true for the case $\Gamma \lsi \Gamma_H$ and $T_R>T^*$
(for details see \cite{massimo}). In order to compute the amplitude and
the spectrum of the baryon number fluctuations we will also suppose
that the Fourier modes of the magnetic fields are stochastically
distributed, leading to a  rotationally
and parity invariant two-point function
\begin{equation}
G_{ij}(r) = \langle H_{i}(\vec{x})
H_{j}(\vec{x}+\vec{r})\rangle~~,
\label{1}
\end{equation}
where $\langle...\rangle$ denotes an ensemble average. In this case
$\langle \delta\left(\frac{n_B}{s}\right)(\vec{x},t)\rangle=0$
\cite{foot1}, but
\begin{equation}
\Delta(r,t_{c}) = \sqrt{\langle\delta
\left(\frac{n_{B}}{s}\right)(\vec{x}, t_{c})
\delta\left(\frac{n_{B}}{s}\right)(\vec{x}+\vec{r},
t_{c})\rangle} \neq 0~.
\label{13}
\end{equation}

Using the  transversality of the  magnetic fields it is useful
to write the two-point function of Eq. (\ref{1}) in Fourier space
\be
G_{ij}(k) = k^2 f(k) (\delta_{ij} - \frac{k_{i} k_{j}}{k^2}).~~
\ee
For $f(k)$ a power spectrum (modified by the typical exponential
decay of small scale magnetic fields given by Eq. (\ref{maxwell})) is
assumed
\be
f(k)=\frac{1}{k}\left( \frac{k}{k_{1}}\right)^{- 4 + \epsilon}
\exp{[-2(\frac{k}{k_{\sigma}})^2]}~,
\label{14}
\ee
where $k_\sigma= T\sqrt{\frac{\sigma_c}{M_0}}$, $k_1$ characterizes the
strength of magnetic fields, and $\epsilon$ is the slope of the
spectrum. A physically realistic situation corresponds to the case in
which the Green's functions of the magnetic hypercharge fields decay at
large distance (i. e. $\epsilon> 0$ in Eq. (\ref{14})) and this would
imply either ``blue''( $\epsilon \geq 0$ ) or ``violet''
($\epsilon \gg 1$) energy spectra. The case of ``red'' spectra
($\epsilon < 0$) will then be left out of our discussion. The flat
spectrum corresponds to $\epsilon \ll 1$ and may appear quite naturally in
string cosmological models \cite{string}.

The explicit result for Eq. (\ref{13}) at the EW phase transition
temperature is:
\begin{eqnarray}
&&\Delta(r,t_{c})= \frac{45 n_{f} \alpha'}{\pi^2 N_{eff}
\sigma_0}\frac{M_{0}}{T_{c}}\frac{\xi^{4-\epsilon}C(\epsilon)}
{(rT_{c})^{1+\epsilon}}(1 + O(\lambda))
\nonumber\\
&&C(\epsilon)=\frac{2^{\epsilon-\frac{3}{2}}\Gamma(\frac{\epsilon}{2}) }
{\Gamma(\frac{3-\epsilon}{2})}~\sqrt{\frac{\pi\epsilon(\epsilon+2)}
{(3-\epsilon)}},
\lambda\sim(\frac{\Gamma}{\Gamma_H})^2 (k_{\sigma}r)^{-2\epsilon}
\label{15}
\end{eqnarray}
(where $\Gamma(z)$ is the Euler Gamma function and  $\xi = k_1/T_c$).
For a flat spectrum of magnetic fields ($\epsilon \ll 1$) baryon number
fluctuations may be rather large. For example, if the energy sitting in
the background magnetic field is comparable with the energy density of
the photons, $\langle {\vec{\cal H}}_{Y}^2 \rangle \sim T^4$ then for
the smallest possible scale $r \sim 1/k_\sigma \sim 10^{-9}\times$(EW
horizon $\simeq 3$cm) we get, from Eq. (\ref{15}) $\delta(n_B/s) \sim
\frac{\alpha'}{N_{eff}}\sqrt{\frac{M_0}{\sigma_c}}\sim 10^3$. This
number exceeds considerably the measure of the baryon asymmetry of the
universe $n_B/s \sim 10^{-10}$, thus small size matter-antimatter
domains are possible at the EW scale. At the same time, for even larger
scales (possibly relevant for structure formation), the fluctuations
of Eq. (\ref{15}) are
quite minute (since their amplitude decreases with the distance as
$1/r^{1+\epsilon}$) and may be safely neglected.

We consider now the question whether the fluctuations we found are
able to affect the standard BBN. This depends upon
the scale of fluctuations at $T=T_c$. Short scale fluctuations (well
inside the EW horizon) have dissipated by the
nucleosynthesis time \cite{10} through the combined action of neutrino
inflation and neutron diffusion. Baryon number fluctuations 
affect BBN provided they are sizable enough over the neutron diffusion
scale ($3\times 10^{5} {\rm cm}$) at the onset of nucleosynthesis
($T_{NS} \simeq 100$ Kev) \cite{10}. The neutron diffusion scale, 
blue-shifted to $T_{c}\simeq 100$ GeV, becomes,
$L_{diff}(T_{c}) = 0.3 ~{\rm cm}$.
Taking again the flat spectrum for magnetic fields and assuming that
their energy is $\sim T^4$ we obtain for the baryon number fluctuations
at that scale $\delta(n_B/s) \sim 10^{-5} \gg 10^{-10}$. If magnetic
fields are large enough, domains of matter and antimatter may exist at
the scales 5 orders of magnitude larger than the neutron diffusion
length. Up to our best knowledge, there were no studies of
non-homogeneous BBN with this type of initial conditions. It would be
very interesting to see whether this may change BBN bounds on the
baryon to photon ratio by changing the related predictions of the light
 element abundances. This possible analysis will not be attempted here.
\begin{figure}
\vspace*{0cm}
\epsfxsize = 7 cm
\centerline{\epsffile{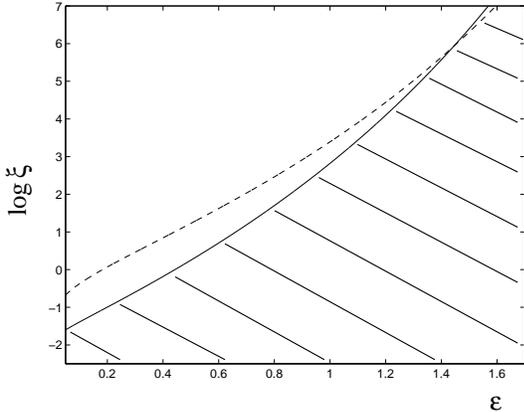}}
\caption[a]{We compare the bound imposed by isotropy requirement at the
BBN scale (Eq. (\ref{19})) with the one imposed by
homogeneity at the same epoch (Eq. (\ref{18})). We see that the
homogeneity requirement (shaded region) is more constraining then the
one of isotropy. The plot is in the case $\Omega_B h^2_{100}=0.01$,
$h_{100}= 0.6$, $\sigma_c/T_c = 70$.}
\label{FIGU}
\end{figure}

A more conservative attitude is to derive bounds on the magnetic fields
from the requirement that homogeneous BBN is not spoiled, i.e.
\begin{equation}
\Delta(L_{diff},t_{c}) < \frac{n_{B}}{s}~.
\label{17}
\end{equation}
In terms of $\xi$ and $\epsilon$ (which completely define our
stochastic magnetic background) the bound (\ref{17}) becomes
\begin{equation}
\log{\xi} < \frac{\log{\frac{\sigma_c}{T_{c}}} - 6.26  +
\frac{1}{2}\log{\epsilon}
+ 14.88~\epsilon +\log{[\Omega_{B} h^2_{100}]}}{4 - \epsilon}~.
\label{18}
\end{equation}
We plot it in {\bf Fig. \ref{FIGU}} for a typical choice of the
parameters and $0.05 \lsi\epsilon \lsi 1.6$ (for $\sigma_{c}/T_{c}\simeq
70-100$, $\Omega_{B}h^2_{100}\simeq 0.1-0.01$, $0.4< h_{100}<1$ this
bound does not change significantly). The bound (\ref{18}) is quite
strong for blue spectra (i. e. $0<\epsilon< 1$). For violet spectra
($\epsilon >1$ in Eq. (\ref{15})) the  fluctuations turns out
to be parametrically smaller than $(n_{B}/s)$ at the neutron diffusion
scale and then practically unconstrained by BBN.

The bounds on ordinary magnetic fields at the nucleosynthesis epoch do
also apply in our case. For example in \cite{3} was obtained that in
order to be compatible with the isotropic nucleosynthesis
$H_{N}(t_{NS}) < 1\times~10^{11}~{\rm Gauss}$ at a temperature $T_{NS}
\sim 10^{9} ~^0{\rm K}$ which implies, at $T_{c}$ , $H_{N}(t_{c}) <
1.34~\times 10^{23}~{\rm Gauss}$. By now comparing the magnetic
hypercharge density with this bound we get a further condition in our
exclusion plot, namely:
\begin{eqnarray}
\log{\xi} < \frac{(11.30 -\frac{1}{2}\log{\frac{\sigma_{c}}{T_{c}}})\epsilon
+\log{\epsilon} - 0.2}{4 - \epsilon}~.
\label{19}
\end{eqnarray}
This condition is reported in {\bf Fig. \ref{FIGU}} (upper curve) and
compared with the one of Eq. (\ref{18}) (lower curve). According to
{\bf Fig. \ref{FIGU}} Eq. (\ref{19}) could be satisfied without
satisfying Eq. (\ref{18}) for $\epsilon \lsi 1.4$. This implies that the
bound we derived is more constraining (by two orders of magnitude for
magnetic field at $\epsilon \ll 1$) than the bounds reported in
\cite{3}.

We thank J. Cline, M. Joyce, H. Kurki-Suonio, and G. Veneziano for
helpful discussions and comments.


\begin{thebibliography}{99}
\bibitem{foot1}If the spectrum is dominated by parity non-invariant
configurations then, in general,  $\langle
\delta\left(\frac{n_B}{s}\right)(\vec{x},t)\rangle\neq0$. Examples of
this types of configurations can be found in \cite{joyce}. If this is
the case, the baryon asymmetry of the universe can be a result of decay
of Chern-Simons condensate \cite{ms}.
\end{thebibliography}
\end{document}